\documentclass[11pt]{article}
\usepackage[english]{babel}
\usepackage{amssymb,amsmath,mathrsfs}
\usepackage{cite}

\newtheorem{remark}{Remark}

\newcommand{\hh}{\mathscr{H}}
\newcommand{\hs}{\mathcal{B}_2(\hh)}
\newcommand{\trc}{\mathcal{B}_1(\hh)}
\newcommand{\bsp}{\mathcal{J}}

\newcommand{\tr}{\mathrm{tr}}

\newcommand{\convo}{\circledcirc}
\newcommand{\fac}{{\frac{1}{2\pi}}}
\newcommand{\intr}{\int_{\mathbb{R}}}
\newcommand{\dx}{\;\mathrm{d}x\hspace{0.4mm}}
\newcommand{\dqdp}{\;\mathrm{d}q\,\mathrm{d}p\hspace{0.4mm}}

\newcommand{\dom}{\mathrm{Dom}}
\newcommand{\sopa}{\mathfrak{A}}

\newcommand{\wigu}{\mathcal{V}}
\newcommand{\ran}{\mathrm{Ran}\hspace{0.3mm}}
\newcommand{\subsp}{\mathfrak{S}(\wigu)}
\newcommand{\subspo}{\mathfrak{S}_0(\wigu)}
\newcommand{\subspw}{\mathfrak{S}(\wign)}
\newcommand{\qp}{{(q,p)}}
\newcommand{\lr}{\mathrm{L}^2(\mathbb{R})}
\newcommand{\lrr}{\mathrm{L}^2(\mathbb{R}\times\mathbb{R})}
\newcommand{\disp}{\exp\!\left(\ima(p\hspace{0.5mm}\hq-q\hspace{0.3mm}\hp)\right)}
\newcommand{\hq}{{\hat{q}}}
\newcommand{\hp}{{\hat{p}}}
\newcommand{\tq}{\tilde{q}}
\newcommand{\tp}{\tilde{p}}
\newcommand{\wign}{\mathcal{W}}
\newcommand{\fs}{\mathcal{F}_{\hspace{-0.6mm}\mbox{\rm \tiny sp}}^{\phantom{x}}}
\newcommand{\fsy}{\mathcal{F}_{\hspace{-0.6mm}\mbox{\rm \tiny sp}}}
\newcommand{\intrr}{\int_{\mathbb{R}\times\mathbb{R}}}
\newcommand{\lurr}{\mathrm{L}^1(\mathbb{R}\times\mathbb{R})}
\newcommand{\shift}{\mathcal{S}}

\newcommand{\ru}{\ran(\wigu)}

\newcommand{\ddu}{d_U^{\phantom{1}}}
\newcommand{\ee}{\mathrm{e}}

\newcommand{\invo}{\hspace{0.4mm}\mathfrak{J}\hspace{0.5mm}}
\newcommand{\sinvo}{\hspace{0.2mm}\mathsf{J}\hspace{0.3mm}}

\newcommand{\starp}{\star}
\newcommand{\Hstar}{\mathrm{H}^\ast\hspace{-0.5mm}}
\newcommand{\gm}{\circledast}
\newcommand{\rr}{{\mathbb{R}\times\mathbb{R}}}

\newcommand{\tre}{\hspace{0.3mm}}
\newcommand{\quattro}{\hspace{0.4mm}}
\newcommand{\cinque}{\hspace{0.5mm}}
\newcommand{\sei}{\hspace{0.6mm}}
\newcommand{\sette}{\hspace{0.7mm}}
\newcommand{\otto}{\hspace{0.8mm}}
\newcommand{\nove}{\hspace{0.9mm}}
\newcommand{\dieci}{\hspace{1mm}}

\newcommand{\err}{\mathsf{R}}
\newcommand{\elle}{\mathsf{L}}

\newcommand{\errep}{\mathbb{R}^{\mbox{\tiny $+$}}}

\newcommand{\erre}{\mathbb{R}}

\newcommand{\ccc}{\mathbb{C}}

\newcommand{\de}{\mathrm{d}}
\newcommand{\eee}{\mathrm{e}}

\newcommand{\opa}{\hat{A}}

\newcommand{\cz}{\mathsf{C}_0}
\newcommand{\czc}{\mathsf{C}_{\mathrm{c}}}

\newcommand{\hrho}{\hat{\rho}}

\newcommand{\mut}{\mu_t}
\newcommand{\mus}{\mu_s}
\newcommand{\must}{\mu_{t+s}}
\newcommand{\cmut}{\check{\mu}_t}
\newcommand{\tmut}{\widetilde{\mu}_t}

\newcommand{\ima}{\mathrm{i}}

\newcommand{\two}{\mathcal{T}}

\newcommand{\intG}{\int_G}

\newcommand{\rrr}{\mathbb{R}}

\newcommand{\urep}{{U\hspace{-0.5mm}\vee\hspace{-0.5mm} U}}
\newcommand{\rep}{{\underline{U\hspace{-0.5mm}\vee\hspace{-0.5mm} U\hspace{-0.8mm}}\hspace{0.8mm}}}

\newcommand{\unir}{\mathfrak{V}}
\newcommand{\cunir}{\check{\mathfrak{V}}}
\newcommand{\unimt}{{\mut[\unir]}}
\newcommand{\unicmt}{{\cmut[\unir]}}
\newcommand{\cunimt}{{\mut[\cunir]}}
\newcommand{\unimo}{{\mu_0[\unir]}}
\newcommand{\unims}{{\mus[\unir]}}
\newcommand{\unimst}{{\must[\unir]}}

\newcommand{\unimu}{{\mut[\urep]}}
\newcommand{\extunimu}{{\mut[\rep]}}

\newcommand{\emme}{\mathsf{m}\hspace{0.3mm}}
\newcommand{\modu}{\Delta_G}
\newcommand{\ldg}{\mathrm{L}^2(G)}

\newcommand{\twoside}{\mathcal{T}_{\emme}\hspace{-0.2mm}}
\newcommand{\temme}{\breve{\mathsf{m}}\hspace{0.3mm}}
\newcommand{\smg}{\mathfrak{T}_t^{\hspace{0.2mm}\emme}}
\newcommand{\smgfw}{\mathfrak{T}_t^{\mbox{\tiny $\wigu$}}}
\newcommand{\smgw}{\mathfrak{T}_t^{\mbox{\tiny $\wign$}}}
\newcommand{\twomt}{{\mut[\twoside]}}

\begin{document}

\title{Operators versus functions: from quantum dynamical semigroups to tomographic semigroups}

\author{Paolo Aniello   \vspace{2mm}\\
\small \it    Dipartimento di Fisica dell'Universit\`a di Napoli ``Federico II'' and INFN -- Sezione di Napoli,  \\
\small \it    Complesso Universitario di Monte S.\ Angelo, via Cintia, 80126 Napoli, Italy}

\date{}

\maketitle

\begin{abstract}
\noindent   Quantum mechanics can be formulated in terms of phase-space functions, according to Wigner's approach.
A generalization of this approach consists in
replacing the density operators of the standard formulation with suitable functions,
the so-called generalized Wigner functions or (group-covariant) tomograms,
obtained by means of group-theoretical methods.
A typical problem arising in this context is to express
the evolution of a quantum system in terms of tomograms.
In the case of a (suitable) open quantum system, the dynamics can
be described by means of a quantum dynamical semigroup `in disguise',
namely, by a semigroup of operators acting on tomograms rather than on density operators.
We focus on a special class of quantum dynamical semigroups, the twirling semigroups,
that have interesting applications, e.g., in quantum information science.
The `disguised counterparts' of the twirling semigroups, i.e.,
the corresponding semigroups acting on tomograms,
form a class of semigroups of operators that we call \emph{tomographic semigroups}.
We show that the twirling semigroups and the tomographic semigroups
can be encompassed in a unique theoretical framework,
a class of semigroups of operators including also the probability semigroups of classical probability theory,
so achieving a deeper insight into both the mathematical and the physical aspects of the problem.
\end{abstract}

%%%--------------------------------------------------------------------------------------------------------------------------

\section{Introduction}
\label{intro}

As is well known, quantum mechanics admits a remarkable
formulation in terms of functions living on phase space,
as first recognized by Wigner in his seminal paper~\cite{Wigner}.
In this approach, further developed by Groenewold~\cite{Groenewold} and Moyal~\cite{Moyal},
every pure state $|\psi\rangle \langle\psi |$ --- a rank-one projector,
with $\psi$ denoting a wave function
(of a single spatial degree of freedom, for notational simplicity)
--- is replaced by a phase space function $\varrho_\psi$:
\begin{equation} \label{defiwf}
\hrho_\psi=|\psi\rangle \langle\psi | \ \; \mapsto \ \; \varrho_\psi (q,p) := \fac\intr \eee^{-\ima p x}\,
\psi\!\left(q-\frac{x}{2}\right)^*
\psi\!\left(q+\frac{x}{2}\right)\dx .
\end{equation}
The (real) function $\varrho_\psi$ is usually called the \emph{Wigner function} --- or distribution ---
associated with the pure state $\hrho_\psi$.
More generally~\cite{Hillery}, with a state $\hrho$ (a density operator)
and with an observable $\hat{A}$ one can suitably associate real
functions $\varrho$ and $\mathcal{A}$, respectively, in such a way that
\begin{equation}
\langle \hat{A}\rangle_{\hrho} =  \tr (\hat{A}\,\hrho\tre) = \intrr \mathcal{A}(q,p)\, \varrho(q,p)
\dqdp ;
\end{equation}
namely, the expectation value $\langle \hat{A}\rangle_{\hrho}$
of the observable $\hat{A}$ in the state $\hrho$
can be represented by a `formally classical' expression
(the function $\varrho$ may assume negative values
so that, in general, it cannot be regarded as a genuine probability distribution).

This intriguing approach has several remarkable applications, ranging from
the study of the classical limit of quantum mechanics~\cite{Landsman}
to quantum optics~\cite{Schleich}, and it can be generalized by replacing
the usual Hilbert space operators, states and observables,
of the standard formulation of quantum mechanics with suitable
functions --- not necessarily living on a standard phase space
(a symplectic vector space, a symplectic manifold) ---
obtained by means of various techniques sometimes collected
under the generic name of \emph{dequantization}, a procedure
which can be regarded as the `reverse arrow' of quantization~\cite{Ali-Englis}.
Specifically, Wigner's correspondence --- operator $\mapsto$ function --- turns out to be
the reverse arrow of Weyl's quantization prescription~\cite{Weyl}; see~\cite{Folland} for
the technical details.
Among all known (quantization-) dequantization schemes, a central role is played
by those schemes based on group-theoretical methods~\cite{Ali,AnielloFT,AnielloSP},
and quantization-dequantization {\it \`a la} Weyl-Wigner is no exception, as it will
be briefly recalled later on.
In this framework, the functions corresponding to quantum-mechanical states
--- the so-called generalized Wigner functions or (group-covariant quantum) \emph{tomograms} ---
live on the symmetry group of a quantum system, or on an homogeneous space of that group,
and the symmetry action of the group on tomograms admits a simple expression.

A natural problem that arises in this general approach
is to express the temporal evolution ---
or the evolution equation --- of a quantum system
in terms of the dequantized objects, i.e., of tomograms.
In the case of a (suitable) \emph{open} quantum system, the temporal evolution
can be described by means of a quantum dynamical semigroup `in disguise', namely,
by a semigroup of operators acting on quantum tomograms in such a way to mimic the action
of a \emph{standard} quantum dynamical semigroup
--- i.e., a completely positive trace-preserving semigroup of operators~\cite{Holevo, Breuer} ---
on states.

In the present paper, we will focus, in particular, on a class of quantum dynamical semigroups
--- the so-called \emph{twirling semigroups} ---
with interesting applications~\cite{Aniello1,Aniello2,Aniello3}.
Each of these semigroups of operators is associated with a pair formed
by a representation of a group and by a convolution
semigroup of probability measures on that group.
Quantum dynamical semigroups of this form were first
considered by Kossakowski~\cite{Kossakowski}, in the early times of the theory of
open quantum systems. The `disguised counterparts' of the twirling semigroups
--- acting on tomograms ---
form a class of semigroups of operators the we call, for rather obvious reasons,
\emph{tomographic semigroups}~\cite{Aniello2,Aniello3}.

We will show that the twirling semigroups
and the tomographic semigroups can be
encompassed in a unique theoretical framework
--- a large class of semigroups of operators
including also the probability semigroups of
classical probability theory --- the
\emph{randomly generated semigroups}.
This unifying theoretical framework
allows one to achieve a deeper insight into both
the mathematical and the physical aspects of the topic.

As a remarkable example, we will study the case where the underlying symmetry group
is the group of translations on phase space.
In this case, the quantum tomograms are directly related to
the Wigner functions, and the explicit form of tomographic semigroups reveals
an interesting relation with classical Brownian motion.

The paper is organized as follows. In sect.~{\ref{framework}},
we introduce the general class of randomly generated semigroups.
Next, in sect.~{\ref{tomograms}},
we describe the special class of dequantized objects (the group-covariant tomograms) that we consider.
This class encompasses, in particular, the standard Wigner functions. We then
derive the general form of tomographic semigroups --- see sect.~{\ref{fromto}}
--- and we consider, in particular, the case of the temporal evolution of Wigner functions.
Finally, in sect.~{\ref{conclusions}}, a few conclusions are drawn.

%%%-------------------------------------------------------------------------------------------------------------------------

\section{A unifying framework: the randomly generated semigroups}
\label{framework}

As anticipated, we will first consider a general class of semigroups
of operators that contains, in particular, both the quantum dynamical semigroups
we are interested in and their `disguised counterparts'.
For the technical details concerning probability theory on groups
(convolution semigroups of measures, probability semigroups, L\'evy-Kintchine formula),
we send the reader to the classical references~\cite{Grenander,Heyer}.

Let $G$ be a \emph{locally compact group}; in particular, it may be a \emph{Lie group}
--- e.g., the symmetry group related to some physical model ---
like the group of translations on phase space, the rotation group,
a (finite-dimensional) unitary group, or the Galilei group.
Suppose that $G$ is endowed with a \emph{convolution semigroup of measures}, i.e., a
family $\{\mut\}_{t\in\errep}$ of probability measures on $G$ such that
\begin{equation}
\mut\convo\mu_{s} = \mu_{t+s},\ \ \ t,s\ge0,
\end{equation}
with $\mut\convo\mu_{s}$ denoting the convolution of the measure $\mut$ with the
measure $\mu_s$, and
\begin{equation}
\lim_{t\downarrow 0} \mut = \delta,
 \ \ \ \delta\equiv\delta_e \ \; \mbox{(Dirac measure at the identity $e$ in $G$)} .
\end{equation}
Let, moreover, $\unir$ be a (weakly continuous)
representation, or \emph{anti}representation, of $G$ in
a real or complex Banach space $\bsp$
--- in the case of an antirepresentation, of course, we will have that
$\unir(g_1 g_2)=\unir(g_2)\sei \unir(g_1)$, for all $g_1, g_2\in G$.
The set $\{\unimt\}_{t\in\errep}$, with $\unimt\colon\bsp\rightarrow\bsp$ denoting the bounded linear operator
defined by
\begin{equation} \label{rgs}
\unimt \sei \Psi := \int_G  \unir(g)\sei \Psi\; \de \mut(g),\ \ \ \forall\quattro\Psi\in\bsp,
\end{equation}
is a \emph{semigroup of operators} (for a rigorous definition of the operator $\unimt$ see~{\cite{Aniello2}}); i.e.,
\begin{enumerate}
\item $\unimt\cinque \unims = \unimst$, $t,s\ge 0$ \ (one-parameter semigroup property);

\item $\unimo = I$ \ ($I$ denoting the identity operator);

\item $\lim_{t\downarrow 0}\|\unimt\cinque\Psi - \Psi\|=0$,
$\forall\quattro\Psi\in\bsp$ \  (strong right continuity at $t=0$).

\end{enumerate}
We call a semigroup of operators of the form~{(\ref{rgs})}
a \emph{randomly generated semigroup}~{\cite{Aniello2,Aniello3}},
associated with the pair $(\unir,\{\mut\}_{t\in\errep})$.

\begin{remark}
{\rm
Being a semigroup of operators, $\{\unimt\}_{t\in\errep}$
is completely characterized by its \emph{infinitesimal generator},
namely, by the closed linear operator $\sopa$ in $\bsp$ defined by
\begin{equation}
\dom(\sopa):=\Big\{\Psi\in\bsp\colon \exists\cinque
\lim_{t\downarrow 0}
t^{-1}\big(\unimt\cinque\Psi-\Psi\big)\Big\},\
\sopa\cinque\Psi := \lim_{t\downarrow 0}
t^{-1}\big(\unimt\cinque\Psi-\Psi\big),\
\Psi\in\dom(\sopa).
\end{equation}
}
\end{remark}

\begin{remark}
{\rm
It is worth noting that the set of the randomly generated semigroups associated with
group representations coincides with the whole set of randomly generated semigroups;
namely, with the randomly generated semigroups associated with
either representations or antirepresentations. Indeed, given a representation (alternatively, an
antirepresentation) $\unir$ of $G$, and denoting by $\cunir$ the related \emph{anti}representation (respectively,
representation) defined by $\cunir(g):=\unir(g^{-1})$, we have
\begin{equation}
\cunimt = \unicmt,
\end{equation}
where $\cmut$ is the \emph{adjoint} of the measure $\mut$, namely, the probability measure
determined by $\int_G  f(g)\; \de \cmut(g)=\int_G  f(g^{-1})\; \de \mut(g)$,
for every $f$ belonging to the space $\czc(G;\erre)$
of continuous $\erre$-valued functions on $G$ with compact support
(the set $\{\cmut\}_{t\in\errep}$ is a convolution semigroup of measures too).
}
\end{remark}

The class of randomly generated semigroups contains, in particular, the following
remarkable subclasses:
\begin{itemize}

\item The \emph{probability semigroups}, that describe the statistical
properties of Brownian motion~\cite{Aniello1,Ito,Nelson}: in this case,
$\unir(g)=\err_g$ or $\unir(g)=\elle_g$, where $\err_g$, $\elle_g$ are the left and right
translation operators, respectively, acting in the Banach
space $\cz(G;\erre)$ of continuous functions on $G$ vanishing
at infinity, i.e.,
\begin{equation}
\err_g\cinque f := f((\cdot)g),\
\elle_g\cinque f := f(g^{-1}(\cdot)),\ f\in \cz(G;\erre).
\end{equation}
In the case where $G$ is a Lie group, the infinitesimal
generators of these semigroups of operators are given by
the L\'evy-Kintchine formula~\cite{Aniello1,Grenander,Heyer}.

\item The \emph{twirling semigroups}, i.e., certain quantum dynamical semigroups
introduced by Kossakowski~\cite{Kossakowski} during the pioneering times of the
theory of open quantum systems: in this case, given a unitary (or, in general,
a projective) representation $U$ of $G$, in a separable complex Hilbert space $\hh$,
and denoting by $\trc$ the Banach space of trace class operators in $\hh$,
we have:
\begin{equation} \label{simac}
\unir(g)\sei\hrho = \urep(g)\sei\hrho := U(g)\sei \hrho\otto U(g)^\ast,\ \ \
\hrho\in\trc;
\end{equation}
namely, $\unir$ is nothing but the standard symmetry action of $G$
on quantum states.
Classical examples of twirling semigroups are the quantum dynamical
semigroups that model a finite-dimensional quantum system
either coupled to an infinite free boson bath
with Gaussian time correlation functions~\cite{Go-Kos},
or in the limit of singular coupling to a
reservoir at infinite temperature~\cite{Fri-Go}. The infinitesimal
generators of the twirling semigroups have been studied
in detail recently~\cite{Aniello1}.

\item The \emph{tomographic semigroups}: in this case, the representation $\unir$
is given by the symmetry action of the group $G$ on the group-covariant tomograms
associated with the representation $U$ of relation~{(\ref{simac})};
the explicit form of this action will be given in sect.~{\ref{fromto}}.

\end{itemize}

As already mentioned, the statistical properties of Brownian motion
are described in terms of convolution semigroups of measures.
Therefore, definition~{(\ref{rgs})} establishes a remarkable link between
Brownian motion on groups and semigroups of operators.

%%%------------------------------------------------------------------------------------------------------------------------

\section{Group-covariant quantum tomograms and Wigner functions}
\label{tomograms}

Although not immediately evident from its usual definition --- see~{(\ref{defiwf})} ---
the Wigner function associated with a pure state has a group-theoretical
content that allows one to extend this definition to a larger class
of objects: the generalized Wigner functions or (group-covariant) tomograms.

Let us first briefly sketch the mathematical framework;
for the details, see~\cite{Ali,Aniello-sdp,Aniello-sipr,Aniello-ext}, and references therein.
Consider a \emph{square integrable} (irreducible) projective representation $U$
of a locally compact group $G$ in a separable
complex Hilbert space $\hh$. Examples of groups admitting representations of this kind
are the group of translations on phase space --- see below ---
and the affine group (whose square integrable unitary representations
are involved in wavelet theory~\cite{Ali,Daubechies}), and
\emph{all} the irreducible unitary representations
of compact groups are square integrable.
By means of such a representation, one can construct a map
\begin{equation}
\wigu \colon\hs\rightarrow\ldg
\end{equation}
--- with $\hs$, $\ldg$ denoting the Hilbert spaces, respectively,
of Hilbert-Schmidt operators in $\hh$,
and of square integrable, with respect to
the left Haar measure, $\ccc$-valued functions on $G$ ---
called the \emph{generalized Wigner map}, or \emph{Wigner transform},
generated by the square integrable representation $U$.
The map $\wigu$ is a (linear) \emph{isometry}, and we will
denote its range by $\ru$. It turns out that $\ru$ depends
only on the unitary equivalence class of the representation
$U$. If the group $G$ is \emph{unimodular} (e.g., a compact group), then the isometry $\wigu$
maps a trace class operator $\hrho\in\trc$
to the function
\begin{equation} \label{nonformal}
\big(\wigu\tre\hrho\big)(g)=d_U^{-1}\,\tr(U(g)^\ast\hrho) ,
\end{equation}
with $\ddu>0$ denoting a normalizing constant,
depending on the representation $U$ and, of course,
on the normalization of the Haar measure.
Since trace class operators are dense in $\hs$,
formula~{(\ref{nonformal})} determines the Wigner transform
completely, in the case of an unimodular group, while in the
general case the definition of $\wigu$ is somewhat more complicated
and involves a suitable positive selfadjoint operator associated with $U$,
the so-called `Duflo-Moore operator' (which, in the unimodular case,
reduces to a multiple of the identity of the form $\ddu\hspace{0.4mm}I$,
where $\ddu$ is the positive constant appearing in~{(\ref{nonformal})}).

The symmetry group that gives rise to the \emph{standard} Wigner transform
is the group of translations on phase space.
For notational simplicity, we will consider the special case of the
$(1+1)$-dimensional the additive group $\rrr\times\rrr$
(a single spatial degree of freedom and the associated momentum),
but the extension to the $(n+n)$-dimensional case is straightforward.
The irreducible unitary representations of this (abelian) group
are obviously one-dimensional; hence, they are not relevant
from the point of view of quantum mechanics. Therefore,
one is lead to consider a suitable class of
irreducible \emph{projective} representations
that are labeled by a nonzero real number whose (absolute) value
can be interpreted as Planck's constant $\hbar$; see, e.g.,~{\cite{Emch,AnielloWS}}.
Setting $\hbar=1$, one has the projective representation
\begin{equation} \label{Weyl-sys}
\rrr\times\rrr\ni\qp\mapsto U\qp:=\disp
\end{equation}
--- where $\hq$, $\hp$ are the standard position and momentum operators in $\lr$ ---
the so-called \emph{Weyl system} (in quantum optics,
$U\qp$ is often called \emph{displacement operator}~{\cite{Schleich,Aniello-coher}).
The fact that we are dealing with a \emph{projective} representation
means, in this case, that
\begin{equation}
U(q + \tq, p + \tp)= \emme(q,p\hspace{0.6mm};\tq,\tp)\sette U(q,p)\cinque U(\tq, \tp),
\end{equation}
where the \emph{multiplier} $\emme$ is of the form
\begin{equation} \label{multip}
\emme(q,p\hspace{0.6mm};\tq,\tp):=\exp\hspace{-0.5mm}\Big(\frac{\ima}{2}(q\tp-p\tq)\Big).
\end{equation}
However, the physically relevant symmetry action is given by the
associated (non-projective) representation $\urep$ of
$\rrr\times\rrr$ in the Banach space $\mathcal{B}_1(\lr)$ --- see~{(\ref{simac})} ---
which extends in a natural way to a unitary representation $\rep$ in the Hilbert space
$\mathcal{B}_2(\lr)$; i.e.,
\begin{equation}
\rep\qp\sei\opa := U\qp\sei \opa\otto U\qp^\ast, \ \ \opa\in\mathcal{B}_2(\lr).
\end{equation}

The Weyl system is a square integrable representation. Hence, one can define
the associated map $\wigu$ by formula~{(\ref{nonformal})} (the additive group $\rrr\times\rrr$ is
obviously unimodular).
It turns out that, in this case, the isometry $\wigu$ that one obtains
is not directly the standard Wigner transform but the so-called
\emph{Fourier-Wigner transform}~\cite{Folland}. The former map
is related to the latter one by the \emph{symplectic Fourier transform}, i.e.,
by the unitary operator $\fs \colon \lrr\rightarrow\lrr$ determined by
\begin{equation}
\big(\fs f\big)(q,p)=\frac{1}{2\pi}\intrr f(q^\prime,p^\prime)\,
\ee^{\ima (qp^\prime - pq^\prime)}\; \de q^\prime \de p^\prime,\ \ \
\forall\hspace{0.4mm} f\in\lurr\cap\lrr  .
\end{equation}
Recall, by the way, that $\fs$ is both unitary and selfadjoint:
\begin{equation}
\fs=\fsy^\ast  ,\ \ \ \fsy^2=I  .
\end{equation}

The symplectic Fourier transform, however, does not play any essential mathematical
or conceptual role, here; it allows one to obtain the usual quantization-dequantization
rules ({\it \`a la} Weyl-Wigner) for position and momentum. Therefore, one should not
expect to have any analogous of this operator involved in the general
group-theoretical quantization-dequantization framework
(namely, for a generic locally compact group).

Going back to the definition of the map $\wigu$, we first note that
$(2\pi)^{-1}\de q\de p$ is the Haar measure
on $\rrr\times\rrr$ normalized in a such a way that $\ddu =1$ for the Weyl system
$U$ (this fact is a consequence of Moyal's identity, see~\cite{Folland}).
Then, in this case, the generalized Wigner transform (the Fourier-Wigner transform) $\wigu$ is
the isometry from $\mathcal{B}_2(\lr)$ into
$\lrr\equiv\mathrm{L}^2\big(\rrr\times\rrr, (2\pi)^{-1}\de q\de
p\hspace{0.3mm};\ccc\big)$ determined by
\begin{equation} \label{deter}
\big(\wigu\tre \hrho\big)\qp =\tr(U\qp^\ast\hrho) ,\ \
\ \forall\hspace{0.4mm}\hrho\in\mathcal{B}_1(\lr)  .
\end{equation}
As a consequence of a result of Pool~\cite{Pool}, in this case $\ru$ coincides with the whole
Hilbert space $\ldg=\lrr$; otherwise stated, the isometry $\wigu$ is actually a unitary operator.

The map $\wigu$ intertwines the unitary
representation $\rep$ with the representation $\two$ of
$\rrr\times\rrr$ in $\lrr)$ defined by
\begin{equation} \label{formtwo}
\big(\two(q,p)\hspace{0.4mm} f\big)(\tq,\tp)=
\ee^{-\ima(q\tp - p\tq)}\hspace{0.8mm} f(\tq,
\tp)  ,\ \ \ \forall\quattro f\in\lrr  .
\end{equation}
Moreover, $\wigu$ intertwines the standard involution $\invo$ in $\mathcal{B}_2(\lr)$
--- namely, the adjoining map $\invo\colon\opa\mapsto\opa^\ast$ --- with the complex
conjugation $\sinvo\colon\lrr\rightarrow\lrr$ (the idempotent antiunitary operator) defined by
\begin{equation} \label{desinvo}
\big(\sinvo f\big)(q,p) = f(-q,-p)^\ast,\ \ \ \forall\hspace{0.4mm}
f\in\lrr  .
\end{equation}
Therefore, one should not expect the image of a selfadjoint operator,
via the map $\wigu$, to be a real function --- as it happens
for a Wigner distribution~\cite{Hillery} --- but rather to satisfy the
relation $f\qp=f(-q,-p)^\ast$.

This fact is not surprising since,
as anticipated, the \emph{standard Wigner transform} --- we will
denote it by $\wign$ --- is the unitary operator obtained composing the
map $\wigu$ with the symplectic Fourier transform:
\begin{equation}
\wign := \fs \sei\wigu\colon \mathcal{B}_2(\lr)\rightarrow \lrr  .
\end{equation}
In the case of a pure state $\hrho_\psi=|\psi\rangle \langle\psi |$, $\|\psi\|=1$,
by a simple calculation one checks that for the function
$\varrho_\psi=\wign\tre\hrho_\psi$ the classical formula~{(\ref{defiwf})}
holds.

Clearly, the unitary operator $\wign$ intertwines the representation
$\rep$ with the unitary representation
$\shift$ of $\rrr\times\rrr$ in $\lrr$
defined by
\begin{equation} \label{shifta}
\shift\qp
:=\hspace{0.3mm}\fs\hspace{0.6mm}\two\qp\hspace{0.7mm}\fs  , \
\ \ \forall\hspace{0.5mm}\qp\in\rrr\times\rrr  .
\end{equation}
It is easy to check that, explicitly, we have:
\begin{equation} \label{shiftb}
\big(\shift\qp\hspace{0.4mm} f\big)(\tq,\tp)=
 f(\tq -q, \tp -p)  ,\ \ \ \forall\hspace{0.4mm}f\in\lrr  .
\end{equation}
Thus, the representation $\shift$ is nothing but the regular
representation of $\rrr\times\rrr$ in $\lrr$;
it acts by simply translating functions on phase space.

\begin{remark}
{\rm
The Hilbert space $\lrr$, regarded as the range of the
Wigner map (the standard Wigner transform, or the map directly associated with
the Weyl system, i.e., the Fourier-Wigner transform), carries a natural
structure of an algebra, endowed with the operation induced by the composition
of operators in $\mathcal{B}_2(\lr)$ (the domain of the unitary operators
$\wigu$ and $\wign$); precisely, it is a $\Hstar$-algebra~\cite{AnielloSP,Rickart}).
This algebra operation is usually called by physicists a `star product' of functions,
see~\cite{Zachos,AnielloFT,AnielloSP} and references therein.
In the case where $\lrr$ is regarded as $\ru$, the star product
is given by the expression
\begin{eqnarray}
\Big(f_1\starp f_2\Big)(q,p)
=  \frac{1}{2\pi}\intrr f_1(q^\prime,p^\prime) \,
f_2(q-q^\prime,
p-p^\prime)\hspace{0.6mm}\exp\hspace{-0.5mm}\Big(\frac{\ima}{2}(qp^\prime-pq^\prime)\Big)
\, \de q^\prime \de p^\prime,
\end{eqnarray}
$\forall\hspace{0.5mm} f_1, f_2\in\lrr$. Therefore, the star product
associated with the Weyl system $U$ is precisely the \emph{twisted
convolution} of functions~{\cite{Folland}}, the space $\big(\lrr, \starp, \sinvo\big)$
is a $\Hstar$-algebra and $\wigu\colon\mathcal{B}_2(\lr)\rightarrow\lrr$ is
an isomorphism of $\Hstar$-algebras.
The unitary operators $\wign$, $\wign^\ast$
(i.e., respectively, the dequantization and the quantization map)
induce a further star product
\begin{equation}
(\cdot)\gm(\cdot)\colon
\lrr\times\lrr\ni(f_1,f_2)\mapsto\hspace{0.3mm}\wign\big(\big(\wign^\ast\hspace{0.3mm}
f_1\big) \big(\wign^\ast\hspace{0.3mm} f_2\big)\big)
\hspace{-0.4mm}\in\lrr  ,
\end{equation}
known as the \emph{twisted product}~{\cite{GLP}}.
Since $\wign = \fs \tre\wigu$ and $\wign^\ast = \wigu^\ast\tre\fs$, we
find that
\begin{equation} \label{defist-bis}
f_1\gm f_2 = \hspace{0.3mm} \fs \big(\big(\fs f_1\big)\starp
\big(\fs f_2\big)\big).
\end{equation}
Using this formula, one obtains that, for every pair of functions
$f_1,f_2$ in $\lurr\cap\lrr$,
\begin{equation} \label{twisted-pr}
\big(f_1\gm f_2\big) (q,p) = \frac{1}{\pi^2}\int_{\rr}
\hspace{-0.7mm} \de q^\prime \de p^\prime\,
\int_{\rr}\hspace{-0.7mm} \de q^{\prime\prime} \de p^{\prime\prime}\
\theta\big(q,p; q^\prime, p^\prime ; q^{\prime\prime},
p^{\prime\prime}\big)\, f_1(q^\prime, p^\prime)\,
f_2(q^{\prime\prime}, p^{\prime\prime})  ,
\end{equation}
where
\begin{equation}
\theta\big(q,p; q^\prime, p^\prime ; q^{\prime\prime},
p^{\prime\prime}\big):=\exp\big(\ima 2(qp^\prime - pq^\prime +
q^\prime p^{\prime\prime} - p^\prime q^{\prime\prime} +
q^{\prime\prime} p - p^{\prime\prime} q)\big).
\end{equation}
The function $\theta\colon (\rr)\times(\rr)\times(\rr)\rightarrow\mathbb{T}$ is the so-called
\emph{Groenewold-Moyal integral kernel}. The symplectic Fourier transform intertwines
the complex conjugation $\sinvo$ defined by~{(\ref{desinvo})}
with the standard complex conjugation in $\lrr$, i.e.,
$\fs \sinvo \hspace{0.8mm} \fs f =f^\ast$. Then $\lrr$, endowed with the twisted
product and with the standard complex conjugation of $\ccc$-valued functions,
is once again a $\Hstar$-algebra~{\cite{Pool}}.
}
\end{remark}

%%%--------------------------------------------------------------------------------------------------------------------------

\section{From quantum dynamical semigroups to tomographic semigroups}
\label{fromto}

In view of the importance, in several applications,
of both the `phase-space approach' to quantum mechanics,
briefly outlined in sect.~{\ref{tomograms}}, and
of the theory of open quantum systems~\cite{Holevo, Breuer}, it is a natural problem
to translate the expressions of quantum dynamical semigroups (and of their
infinitesimal generators) in the language of group-covariant tomograms.

Let $G$ a locally compact group and let $\modu$ be the \emph{modular function} on $G$.
Suppose that $\emme\colon G\times G\rightarrow\mathbb{T}$ --- with $\mathbb{T}$ denoting the
circle group of complex numbers of modulus one --- is a \emph{multiplier} for $G$, i.e., that
\begin{equation}
\emme(g,e)=\emme(e,g)=1  ,\ \ \ \forall\quattro g\in G ,
\end{equation}
and
\begin{equation} \label{multirelat}
\emme(g_1,g_2g_3)\, \emme(g_2,g_3)= \emme(g_1 g_2,g_3)\, \emme(g_1,g_2),
\ \ \ \forall\quattro g_1,g_2,g_3\in G .
\end{equation}
E.g., a multiplier for the group of translations on phase space
is given by~{(\ref{multip})}, the multiplier associated with
the Weyl system. Consider now the map
$\twoside\colon G\rightarrow \mathcal{U}(\ldg)$ --- with $\mathcal{U}(\ldg)$
denoting the unitary group of the Hilbert space $\ldg$ --- defined by
\begin{equation} \label{two-sided}
\big(\twoside(g)\tre f\big)(h) := \modu (g)^{\frac{1}{2}}\hspace{2mm}
\temme(g,h)\hspace{0.8mm} f(g^{-1}h g) ,\ \ \ f\in\ldg,
\end{equation}
where the function $\temme\colon G\times G\rightarrow \mathbb{T}$ is
given by the expression
\begin{equation} \label{exprtemme}
\temme(g,h) := \emme (g,g^{-1} h)^\ast \otto \emme (g^{-1} h,g), \ \ \ \forall\quattro g,h\in G.
\end{equation}
One can prove that the map $\twoside$ is a (strongly continuous) unitary representation~\cite{AnielloSP}.

Therefore, given a convolution semigroup $\{\mut\}_{t\in\errep}$ of measures on $G$, we can define
the randomly generated semigroup associated with the pair $(\twoside,\{\mut\}_{t\in\errep})$, namely, the
semigroup of operators
\begin{equation}
\smg f := \twomt \sei f =\intG  \twoside(g)\tre f \; \de\mut(g),
\ \ \ \forall\quattro  f\in\ldg.
\end{equation}
It is then natural to wonder whether the semigroup of operators
$\{\smg\}_{t\in\errep}$ has a physical meaning.
One can answer to this question at least in the case where
there exists a square integrable projective representation
$U$ of the group $G$, with multiplier $\emme$.

Indeed, in this case it can be shown that the semigroup of operators $\{\smg\}_{t\in\errep}$ is essentially
\emph{the twirling semigroup associated with the pair} $(U,\{\mut\}_{t\in\errep})$~\cite{Aniello1},
\emph{but expressed in terms of the quantum tomograms
associated with} $U$. Thus, it is quite natural to call $\{\smg\}_{t\in\errep}$
the \emph{tomographic semigroup}
associated with the multiplier $\emme$.

More precisely, it turns out that the tomograms form a closed subspace of $\ldg$
--- the range $\ru$ of the Wigner map $\wigu$ associated with $U$ ---
\emph{stable} under the action of both the unitary
representation $\twoside$ and the semigroup of operators $\{\smg\}_{t\in\errep}$.
The range $\ru$ contains a distinct (dense) linear subspace $\subsp$ which is the
image, via the isometry $\wigu$, of the Banach space $\trc$. Moreover,
$\subsp$ contains a convex set $\subspo$ which is the image of the convex set of
density operators in $\hh$ (the physical states). The subspace $\subsp$ and the
convex set $\subspo$ are again stable in the sense mentioned before.
One can show that the representation $\urep$ is related to the representation
$\twoside$ by the intertwining property
\begin{equation}
\twoside (g)\otto\wigu\sei\hrho = \wigu\nove\urep(g)\sei\hrho,
\ \ \ \forall\quattro\hrho\in\trc .
\end{equation}
It follows that, for every $t\ge 0$, we have:
\begin{equation} \label{intewi}
\smg\quattro\wigu\sei\hrho = \wigu\sei\unimu\sei\hrho , \ \ \ \forall\quattro\hrho\in\trc,
\end{equation}
where $\{\unimu\colon\trc\rightarrow\trc\}_{t\in\errep}$ is the twirling semigroup
associated with the pair $(U,\{\mut\}_{t\in\errep})$, i.e.,
the quantum dynamical semigroup determined by
\begin{equation}
\unimu\sei\hrho = \int_G  \urep(g)\sei\hrho\; \de \mut(g) .
\end{equation}
Incidentally, we note that the twirling semigroup $\{\unimu\}_{t\in\errep}$
extends in a natural way to the randomly generated semigroup
$\{\extunimu\colon\hs\rightarrow\hs\}_{t\in\errep}$, for which an intertwining relation
analogous to~{(\ref{intewi})} holds.

As the subspace $\subsp$ is stable under the tomographic semigroup, we can further introduce
a semigroup of operators $\{\smgfw\}_{t\in\errep}$ by setting
\begin{equation}
\smgfw \varrho := \smg \varrho, \ \ \ \forall\quattro \varrho\in\subsp.
\end{equation}
Note that we have denoted by $\varrho$ a generic element of the space $\subsp$ in  order
to highlight the fact that it corresponds, via dequantization, to a trace class operator $\hrho$.
We stress that, whereas the tomographic semigroup $\{\smg\}_{t\in\errep}$ can always be defined
(for every multiplier $\emme$), by virtue of relation~{(\ref{intewi})}
the semigroup of operators $\{\smgfw\}_{t\in\errep}$ is the \emph{bona fide} dequantized version
of the twirling semigroup $\unimu$, whose definition relies on the fact that
$U$ is a square integrable representation; hence, it can be regarded as a quantum dynamical semigroup
`in disguise'. We will then call $\{\smgfw\}_{t\in\errep}$ a \emph{proper tomographic semigroup}.

Let us consider the case where $G$ is the group of
translations on the $(1+1)$-dimensional phase space,
i.e., the additive group $\rrr\times\rrr$. In this case,
we can define \emph{two} (proper) tomographic semigroups ---
one associated with the Fourier-Wigner transform ($\{\smgfw\}_{t\in\errep}$), the other
with the standard Wigner transform ($\{\smgw\}_{t\in\errep}$) --- mutually related by the
symplectic Fourier transform. We will now briefly describe these semigroups of
operators.

To this aim, note that for the function $\temme\colon
(\rrr\times\rrr)\times(\rrr\times\rrr)\rightarrow\mathbb{T}$, defined according to~{(\ref{exprtemme})},
we find the expression
\begin{equation}
\temme(q,p\hspace{0.6mm};\tq,\tp)=\emme(q,p\hspace{0.6mm};\tq - q,\tp -p)^\ast\,
\emme(\tq -q,\tp -p\hspace{0.6mm};q,p)=\exp\hspace{-0.5mm}\big(\hspace{-1.2mm}-\ima(q\tp - p\tq)\big),
\end{equation}
and, therefore, as anticipated in sect.~{\ref{tomograms}} the unitary representation
$\two\equiv\twoside \colon\rrr\times\rrr\rightarrow\mathcal{U}(\lrr)$ has the form~{(\ref{formtwo})}.
Then, the proper tomographic semigroup $\{\smgfw\}_{t\in\errep}$ is given by
\begin{equation}
\big(\smgfw \tre\varrho\big) (\tq,\tp)  =
\intrr \big(\two(q,p)\quattro \varrho\big) (\tq,\tp) \; \de \mut (q,p) =
\varrho(\tq,\tp) \intrr \ee^{\ima(\tq p - \tp q)}  \; \de \mut (q,p).
\end{equation}
Hence, denoting by $\tmut$ the symplectic Fourier transform of the probability measure $\mut$,
we have:
\begin{equation} \label{pomul}
\big(\smgfw \tre\varrho\big) (\tq,\tp)  = \tmut(\tq,\tp)\tre \varrho(\tq,\tp), \ \ \ t\ge 0.
\end{equation}

\begin{remark}
{\rm
The expression~{(\ref{pomul})} of the tomographic semigroup $\{\smgfw\}_{t\in\errep}$
has an interesting interpretation. In fact, according to Bochner's theorem~\cite{Rudin}, the (symplectic)
Fourier transform of a probability measure on $\rrr\times\rrr$ is a \emph{normalized}
continuous positive definite function, i.e.,
a continuous positive definite function $f\colon\rrr\times\rrr\rightarrow\ccc$ such that $f(0,0)=1$.
Hence, the tomographic semigroup $\{\smgfw\}_{t\in\errep}$ acts by (pointwise)
multiplication by a positive definite function.
Considering the fact that the space $\subsp$ where the semigroup of operators acts can be characterized as
the set of continuous KLM-positive definite functions on $\rrr\times\rrr$ --- where the acronym refers
to the fundamental contributions of Kastler~\cite{Kastler}, and of Loupias and Miracle-Sole~\cite{LMS1,LMS2} ---
here we have a nice interplay between the `classical' positive definite functions
and the `quantum' KLM-positive definite functions.
}
\end{remark}

Let us now express the tomographic semigroup in terms of (standard) Wigner functions
--- $\varrho =\wign\tre \hrho = \fs \tre\wigu\tre\hrho$ --- by setting
$\smgw  :=  \fs \dieci \smgfw \sei \fs$, i.e.,
\begin{equation}
\smgw \tre\varrho  =
\fs \intrr \two(q,p)\otto (\fs\cinque \varrho) \; \de \mut (q,p) .
\end{equation}
By relations~{(\ref{shifta})} and~{(\ref{shiftb})} we then find that
\begin{equation} \label{fosmgw}
\big(\smgw \tre\varrho \big) (\tq,\tp) =
\intrr \big(\shift\qp\tre \varrho\big)(\tq,\tp)\; \de \mut (q , p) =
\intrr  \varrho(\tq - q,\tp -p)  \; \de \mut (q , p).
\end{equation}
Thus, as in the case of (classical) Brownian motion in $\erre^3$, we have a semigroup
of operators acting on a function $\varrho$ by taking the convolution of this function
with a probability measure belonging to a convolution semigroup. It should be noted, however,
that the function $\varrho$ in~{(\ref{fosmgw})} lives on phase space (rather than on configuration space)
and is characterized as an element of the space $\subspw=\fs\subsp$ that contains the quantum states.

%%%--------------------------------------------------------------------------------------------------------------------------

\section{Conclusions and perspectives}
\label{conclusions}

A state in classical statistical mechanics is a probability measure
on phase space --- typically, associated with a probability distribution (a function),
the Liouville density ---
and, according to a celebrated result of Bochner, the (symplectic)
Fourier transform of a classical state is a continuous positive definite function.
In the quantum mechanical setting, states are usually realized as density
operators, but an elegant description in terms of functions is still possible by
a suitable dequantization scheme. In particular, the properties of
square integrable representations allow one to map a density operator
to a generalized Wigner function or group-covariant quantum tomogram. These tomograms
are embedded in an algebra of functions, with the algebra operation --- a star product ---
corresponding to the product of operators~\cite{AnielloSP}. In the case of a Weyl system (a square integrable
projective representation of the group of translations on phase space),
a tomogram is given by a Fourier-Wigner distribution
--- a normalized continuous KLM-positive definite function --- or by its symplectic
Fourier transform, namely, by a standard Wigner distribution,
a `quasi-probability distribution'~{\cite{Schleich,Aniello-coher}.

We have explored the idea of re-expressing a remarkable
class of quantum dynamical semigroups
--- the twirling semigroups --- in terms of `phase-space functions',
i.e., of group-covariant quantum tomograms.
The implementation of this idea leads to the notion of
\emph{tomographic semigroup}, a quantum dynamical semigroups `in disguise'.
Both the twirling semigroups and their disguised counterparts
are contained in a larger class of semigroups of operators,
the randomly generated semigroups.
These semigroups of operators are associated with pairs of the type
$(\unir,\{\mut\}_{t\in\errep})$, where $\unir$ is a
representation, or an antirepresentation, of a locally compact group $G$
and $\{\mut\}_{t\in\errep}$
is a convolution semigroup of measures on $G$. Interestingly,
the class of randomly generated semigroups also contains the probability semigroups of
classical probability theory, and a tomographic semigroup acting on Wigner distributions
is formally similar to a probability semigroup associated with the group of translations
on phase space, see~{(\ref{fosmgw})}.

We have chosen to keep the exposition at a mild level of mathematical sophistication
in such a way to skip several technical details and to focus on the main ideas. E.g., we
have omitted any explanation concerning the precise meaning of the integrals of
vector-valued functions. For these details, the reader is referred to~\cite{Aniello1,Aniello2}.

The results outlined in the present paper may be extended in several directions.
In particular, there is ongoing work devoted to the characterization of the
infinitesimal generators of the tomographic semigroups. In this regard, it should be noted
that, in spite of the formal similarity with the probability semigroups, the machinery and the techniques
associated with the classical L\'evy-Kintchine formula cannot be applied directly
in the new setting, and one needs to solve certain nontrivial mathematical issues.

%%%--------------------------------------------------------------------------------------------------------------------------

\end{document}